\documentclass[preprint]{emulateapj-rtx4}

\def\ha{{\rm\,H$\alpha$}}
\def\hb{{\rm\,H$\beta$}}

\def\oiii{{\rm\,[O{\sc iii}]}}
\def\pa{{\rm\,P$\alpha$}}

\def\apjl{ApJL}

\slugcomment{Draft version: April 28, 2015}

\shorttitle{NB-selected \oiii\ emission line galaxies at $z > 3$}
\shortauthors{Suzuki et al.}

\begin{document}


\title{Galaxy formation at $z > 3$ revealed by narrow-band selected \\ \oiii\ emission line galaxies}


\author{Tomoko L. Suzuki\altaffilmark{1,2}, Tadayuki Kodama\altaffilmark{1,2}, Ken-ichi Tadaki\altaffilmark{3}, Masao Hayashi\altaffilmark{2}, Yusei Koyama\altaffilmark{4}, 
Ichi Tanaka\altaffilmark{5},  Yosuke Minowa\altaffilmark{5}, Rhythm Shimakawa\altaffilmark{1,5}, and Moegi Yamamoto\altaffilmark{1,2}}


\altaffiltext{1}{Department of Astronomical Science, The Graduate University for Advanced Studies (SOKENDAI), Mitaka, Tokyo 181-8588, Japan}
\altaffiltext{2}{Optical and Infrared Astronomy Division, National Astronomical Observatory of Japan, Mitaka, Tokyo 181-8588, Japan}
\altaffiltext{3}{Max-Planck-Institut f\"{u}r Extraterrestrische Physik, Giessenbachstrasse, D-85748 Garching, Germany}
\altaffiltext{4}{Institute of Space Astronomical Science, Japan Aerospace Exploration Agency, Sagamihara, Kanagawa 252-5210, Japan}
\altaffiltext{5}{Subaru Telescope, National Astronomical Observatory of Japan, 650 North A'ohoku Place, Hilo, HI 96720, USA}


\begin{abstract}
\noindent
We present the physical properties of \oiii\ emission line galaxies at $z>3$ 
as the tracers of active galaxies at 1Gyr before the peak epoch at $z\sim2$.
We have performed deep narrow-band imaging surveys in the
Subaru/{\it XMM-Newton} Deep Survey Field with MOIRCS on the Subaru Telescope 
and have constructed coherent samples of 34 \oiii\ emitters
at $z=3.2$ and 3.6, as well as 107 \ha\ emitters at $z=2.2$ and 2.5.
We investigate their basic physical quantities,
such as stellar masses, star formation rates (SFRs), and sizes using the
publicly available multi-wavelength data and high resolution images
by the {\it Hubble Space Telescope}. 
The stellar masses and SFRs show a clear correlation known as the
 ``main sequence'' of star-forming galaxies.
 It is found that the location of the main sequence of the \oiii\ emitters at $z=3.2$ and 3.6 
 is almost identical to that of the \ha\ emitters at $z=2.2$ and 2.5.
Also, we investigate their mass--size relation and find that the relation does not change between the two epochs.
When we assume that the star-forming galaxies at $z=3.2$ grow simply {\it along} the same main
sequence down to $z=2.2$, 
galaxies with $M_* = 10^{9}$--$10^{11} {\rm M_\odot}$ increase their stellar masses significantly by a factor of 10--2.
 They {\it climb up} the main sequence, and 
 their star formation rates also increase a lot as their stellar masses grow.
 This indicates that star formation activities of galaxies are accelerated from $z>3$ 
 towards the peak epoch of galaxy formation at $z\sim2$.  
\end{abstract}



\section{Introduction}
The activities of star formation in galaxies and those of active galactic nuclei (AGN)
are very high at $z \sim 1-3$, corresponding to about 8--10 billion years ago (e.g. \citealt{hopkins06}; \citealt{fan04}).
Physical states of galaxies are expected to change dramatically during this epoch, 
and it is critical to investigate galaxy properties in detail at this epoch 
in order to understand physical processes of galaxy formation.
With the near-infrared (NIR) photometric and spectroscopic observations with ground-based and space telescopes, 
studies of high-$z$ galaxies have advanced substantially in recent years, 
and our knowledge of physical states of those galaxies have been expanded significantly by many previous studies 
(e.g. \citealt{erb06a, vandokkum08, kriek09b, forsterschreiber09, wuyts11}).

An example is the discovery of the relationship between stellar mass and star formation rate (SFR) of star-forming galaxies 
both at low and high redshifts.
Many previous studies have revealed that these two quantities show a tight correlation, 
called the ``main sequence'' of star-forming galaxies (e.g. \citealt{elbaz07, noeske07, daddi07, whitaker12, kashino13}).
Using galaxy samples across a wide redshift range, 
the evolution of this relation has been investigated (e.g. \citealt{whitaker12, koyama13, tasca14}).
They have shown that the $M_*$--SFR relation does evolve strongly with redshift at least up to $z\sim2.5$, 
in the sense that SFR increases monotonically with redshift at a given stellar mass.

Another example is the discovery of compact, massive, and quiescent galaxies at $z\sim2$ 
(``red nuggets''; \citealt{daddi05, damjanov09}), 
and their likely progenitors, namely, compact star-forming galaxies at $z\ge2$.
The compact star-forming galaxies are thought to be formed by gas-rich processes such as mergers or disk instabilities 
which invoke starbursts in the central compact regions.
They would then become compact, quiescent galaxies when their star formation activities are quenched \citep{barro13}. 
We can search for such compact star-forming galaxies based on the stellar mass versus size diagram.
The redshift evolution of the mass--size relation is crucial for identifying the evolutionary
paths from star-forming galaxies to quiescent galaxies, and it has been investigated up to $z\sim3$.
It is found that the sizes of early-type galaxies depend more strongly on their stellar
masses as compared to late-type galaxies 
and that the average size evolution at a fixed stellar mass of late-type galaxies is very slow, 
while the average size of the early-type galaxies increases rapidly with the cosmic time 
(e.g. \citealt{shen03}; \citealt{vanderwel14}).

The studies of physical properties of galaxies are now being expanded to 
more distant Universe beyond the highest peak of the star formation activities at $z\sim2-3$.
The epoch of $z\sim 3-3.7$, corresponding to about 1--2\ Gyr before the peak epoch, 
is especially crucial to reveal how galaxy formation activities are activated towards its peak.
At $z>2.5$, the ultraviolet (UV) light is often used to construct star-forming galaxy samples.
Using the UV-selected galaxies, such as the Lyman Break Galaxies (LBGs), 
the redshift evolution of star formation activities have been investigated 
(e.g. \citealt{stark09, gonzalez10, reddy12, stark13, tasca14}). 
\citet{stark13} investigate the evolution of the specific star formation rate (${\rm sSFR=SFR/M_*}$) 
for the spectroscopically confirmed LBG sample at $3.8<z<5$.
Their result indicates that the sSFR at the fixed stellar mass increases by a factor of 5 from $z\sim2$ to $z\sim7$.
Based on the $i$-band selected and spectroscopically confirmed galaxy sample, 
\citet{tasca14} investigate the evolution of the $M_*$--SFR and $M_*$--sSFR relations up to $z\sim5$.
Their conclusion is that the sSFR increases very slowly from $z \sim 3$ up to $z \sim 5$.
These two galaxy samples have some small differences in the sample selection (see the papers for details). 
Moreover, both of the samples are likely to be biased to less dusty star-forming galaxies 
since the samples are selected at the rest-frame far UV.
Up to $z\sim2.5$, the \ha\ emission line is a good tracer of star formation 
activity, due to its lower sensitivity to the dust extinction as compared to the UV light.
However, since the \ha\ line comes to longer wavelength than {\it K}-band at $z>3$, 
we are no longer able to use the \ha\ line as the tracer of star-forming galaxies at $z>3$
in the observations with ground-based telescopes.
Recent NIR spectroscopic observations have revealed 
that the high-$z$ star-forming galaxies show strong \oiii\ emission lines   
(e.g. \citealt{masters14, holden14, steidel14, shimakawa14}).
Such a strong \oiii\ emission line indicates extreme interstellar medium (ISM) conditions of high-$z$ star-forming galaxies.
\citet{nakajima14} have shown that the lower metallicities and higher ionization parameters contribute to 
their extreme ISM conditions.
Therefore, we argue that the \oiii\ emission line is one of the best tracers of star-forming galaxies at $z>3$.
We should note however that the \oiii\ emission line originates from ionized regions 
not only by hot young massive stars but also by AGNs.

We have conducted a systematic narrow-band imaging survey with Subaru Prime Focus Camera (Suprime-Cam) and 
Multi-Object InfraRed Camera and Spectrograph (MOIRCS) on the Subaru telescope.
The project is called {\it ``MAHALO-Subaru''} (MApping HAlpha and Lines of Oxygen with Subaru;  \citealt{kodama13}).
The imaging observations with narrow-band filters allow us to obtain emission line galaxies in particular redshift slices.
By targeting high density regions, such as clusters or protoclusters, as well as the lower density
blank fields, we have constructed a coherent NB-selected galaxy sample across various
environments and cosmic times.
\citet{tadaki13} carried out NB imaging observations with MOIRCS at the Subaru/{\it XMM-Newton}
Deep survey Field (SXDF; \citealt{furusawa08}), and have constructed the \ha\ emitter
sample at $z$=2.2 and 2.5.
They found that the \ha\ emitters constitute the star-forming main sequence 
and that galaxies with high SFRs located above the main sequence
tend to be dustier.
They also investigated the structural properties of the \ha\ emitters using 
the {\it Hubble Space Telescope (HST)} images \citep{tadaki14}.
They found two classes of objects; 
star-forming galaxies with extended disks 
and massive and compact star-forming galaxies. 
The latter class of objects are likely to evolve to the red nuggets seen at similar or lower redshifts 
after quenching their star formation activities 
and eventually to massive elliptical galaxies today after significant size growth by minor mergers. 
(e.g. \citealt{vandokkum10})
In order to make such evolutionary links further, 
it is important to explore those possible progenitors 
(massive and compact star-forming galaxies) at higher redshifts, $z>3$, in a systematic way.

In this study, we construct a NB-selected \oiii\ emitter sample at $z>3$ based on the
NB imaging survey data at SXDF taken through the MAHALO-Subaru project \citep{tadaki13} .
We investigate their physical properties to reveal galaxy formation at the epoch
shortly before the highest peak of activities.
In this field, a range of multi-wavelength photometric data and high resolution images
by {\it HST} are available.
Combining those data with our NB imaging data, we investigate stellar masses, SFRs,
and sizes of the \oiii\ emitters.
We will then compare their properties with those of the \ha\ emitters at lower redshifts
($2.2<z<2.5$) in the same field to track the evolution of star-forming galaxies
between the two epochs.

This paper is organized as follows:
In Section 2, we briefly introduce our NB imaging observations and the publicly available observational data 
in SXDF and explain the selection method of emission line galaxies. 
In Section 3, we present the measurements of basic physical quantities of the \oiii\ emitters,
such as stellar masses, SFRs, and sizes, and show the relations between these physical quantities.
We also compare our galaxy sample with the \ha\ emitters at lower redshifts in the same field.
In Section 4, we discuss any possible selection bias arising from our use of the \oiii\ emission
line as the tracer of star-forming galaxies at $z>3$. 
We also discuss the mass growth and star formation histories of galaxies from $z=3.2$ to $z=2.2$ with a simple model.
Finally, we summarize our study in Section 5.
We assume the cosmological parameters of $\Omega_{\mathrm{m}} = 0.27$, $\Omega_{\Lambda} = 0.73$, 
and $H_{\mathrm{0}}=71 \ [\mathrm{km\ s^{-1}\ Mpc^{-1}}]$.
Through this paper, unless otherwise noted, all the magnitudes are given in the AB magnitude
system \citep{oke83}, and the Salpeter initial mass function \citep{salpeter55}
is adopted for the estimations of stellar masses and SFRs.

\section{Observational Data and Sample Selection}
\subsection{Narrow-band Imaging Surveys}
We summarize below our NB imaging observations performed at SXDF,
and further details of the observations and data reduction are described in \citet{tadaki13}.

\begin{figure}[tbp]
\begin{center}
\includegraphics[width=0.25\textwidth, angle=270]{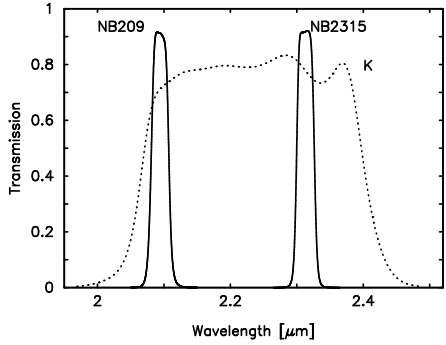}
\caption{Transmission curves of the two narrow-band filters on MOIRCS; NB209 and NB2315 (solid lines), 
and the $K$-band filter on UKIRT/WFCAM (dotted line).
$K$-band data is used to estimate the stellar continuum (Section 2.3).}
\label{figure1}
\end{center}
\end{figure}

The NB imaging at SXDF was carried out with the NIR camera and spectrograph
called MOIRCS \citep{suzuki08} on the Subaru telescope.
MOIRCS is equipped with two HAWAII-2 detectors ($2048 \times 2048$ pixels).
The pixel scale is $0.117$ arcsec/pixel and the field-of-view (FoV) is $4' \times 7'$.
Two NB filters were used, namely,
NB209 ($\lambda_{\rm c} = 2.093\ \mu {\rm m}$, $\rm FWHM = 0.026\ \mu m$) 
and NB2315 ($\rm \lambda_{\rm c} = 2.317\ \mu m$, $\rm FWHM = 0.026\ \mu m$).
The NB209 and NB2315 filters can probe \ha\ emission lines at $z=2.191 \pm 0.019$ and $z=2.525 \pm 0.021$, 
and also \oiii $\lambda 5007$ emission lines at $z=3.174 \pm 0.025$ and $z=3.623 \pm 0.027$, respectively.
Figure \ref{figure1} shows the transmission curves of these two NB filters.
The data were obtained over several observing runs from October to November in 2010 and in September 2011.
The total observed areas were $91\ {\rm arcmin^2}$ and $93\ {\rm arcmin^2}$ for NB209 and NB2315, respectively.
The survey areas are slightly different between the two NB filters, 
but both are overlapped with the CANDELS-UDS field observed by
{\it HST}/ACS and WFC3 (see Figure 2 of \citealt{tadaki13}).
The exposure times were 140--186 minutes, and the seeing sizes were $0.5''-0.7''$ (FWHM). 
The 5$\sigma$ limiting magnitudes with $1.6''$ diameter aperture
were 23.6 mag and 22.88 mag in NB209 and NB2315, respectively.
The observed data were reduced with the MOIRCS imaging pipeline software 
({\tt MCSRED}\footnote[1]{http://www.naoj.org/staff/ichi/MCSRED/mcsred.html}; \citealt{tanaka11}).
The point spread functions (PSF) were smoothed to $0.7''$ when combining all the images.

\subsection{Public Data}
In our survey field, SXDF, the multi-wavelength data from UV to mid-infrared (MIR) are all available.
We use the public photometric catalog provided at the Rainbow Database\footnote[2]{https://arcoiris.ucolick.org/Rainbow\_navigator\_public/}
(\citealt{galametz13}). 
It contains $u$-band data from CFHT/Megacam (O.\ Almaini et al. in preparation), 
$B$-, $V$-, $R_{c}$-, $i'$-, and $z'$-band data from Subaru/Sprime-Cam (SXDS; \citealt{furusawa08}), 
$Y$- and $K_s$-band data from VLT/HAWK-I (HUGs; \citealt{fontana14}), 
$J$-, $H$-, and $K$-band data from UKIRT/WFCAM (UKIDSS; \citealt{lawrence07}), 
$3.6 \mu$m, $4.5 \mu$m, $5.8 \mu$m, and $8.0 \mu$m data from {\it Spitzer}/IRAC, 
and $24\ \mu$m data from {\it Spitzer}/MIPS (SpUDS; PI: J. Dunlop and SEDS; \citealt{ashby13}).

Our survey areas are also mostly covered by the {\it HST}/CANDELS \citep{grogin11, koekemoer11}, 
and the photometric catalog and the high resolution images of $V_{F606W}$-, $I_{F814W}$-band from ACS, 
and $J_{F125W}$-, $H_{F160W}$-band from WFC3 are all publicly available.
All the publicly available multi-wavelength data used in this study are summarized in Table \ref{table1}.

\begin{table}[td]
\caption{SXDF Data Set}
\begin{center}
\begin{tabular}{ccc}
\hline
Instrument & Filter & 5$\sigma$ limiting magnitude \\ \hline\hline
CFHT/MegaCam & {\it u} & 27.68 \\
Subaru/Suprime-Cam & {\it B} & 28.38 \\
 & {\it V} & 28.01 \\
 & {\it $R_c$}  & 27.78 \\
 & {\it $i'$} & 27.69 \\
 & {\it $z'$} & 26.67 \\
{\it HST}/ACS &  $F606W$ & 28.49 \\
 & $F814W$ & 28.53 \\
{\it HST}/WFC3 &  $F125W$ & 27.35\\
& $F160W$ & 27.45 \\
VLT/HAWK-I & {\it Y} &  26.73 \\
& {\it $K_s$}  & 25.92 \\ 
UKIRT/WFCAM & {\it J} &  25.63 \\
& {\it H} & 24.76 \\
& {\it K} & 25.39 \\
 {\it Spitzer}/IRAC & 3.6$\mu$m & 24.72 \\
& 4.5$\mu$m & 24.61 \\
& 5.8$\mu$m & 22.30 \\
& 8.0$\mu$m & 22.26 \\
{\it Spitzer}/MIPS & 24$\mu$m & 30-60 $\mu$Jy \\ \hline
\end{tabular}
\end{center}
\label{table1}
\end{table}%

\subsection{Selection of \oiii\ emitters at $z>3$}

\begin{figure*}[ht]
\centering\includegraphics[width=0.3\textwidth, angle=270]{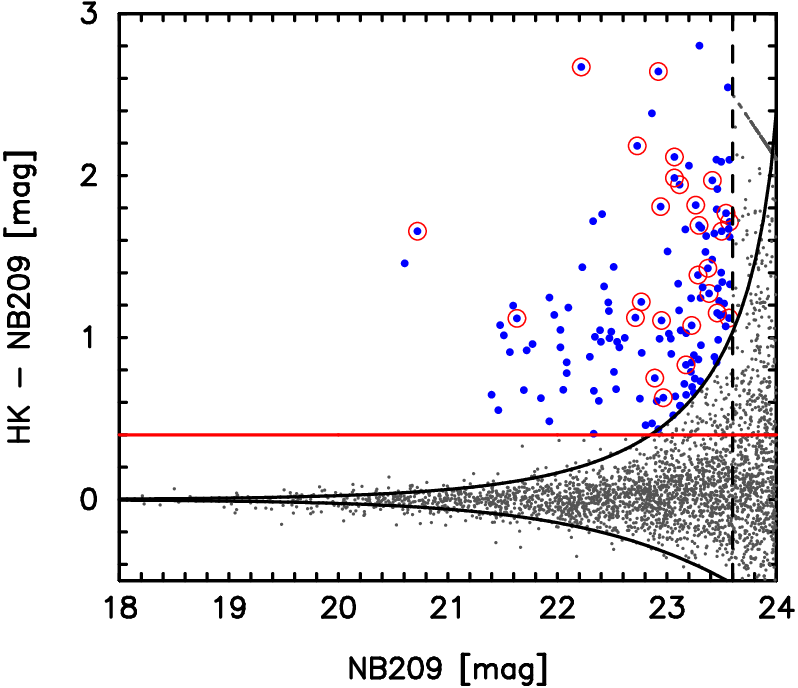}
\hspace{10mm}
\centering\includegraphics[width=0.3\textwidth, angle=270]{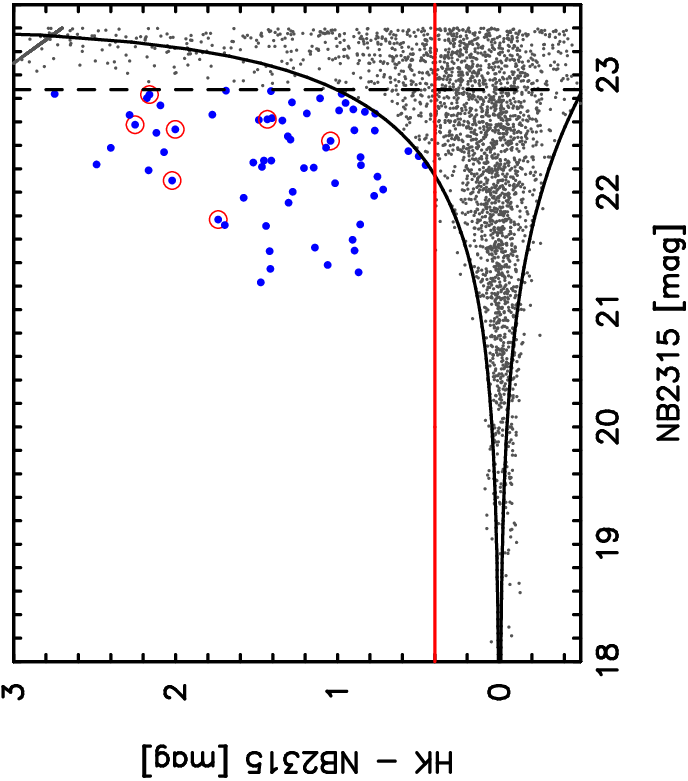}
\caption{The color--magnitude diagrams for NB209 (left) and NB2315 (right). 
Gray dots represent all the sources detected in the NB images, and 
blue filled circles are the sources selected as the NB emitters
with our selection criteria.
Red open circles show the \oiii\ emitters at $z=3.17$ (left) and $z=3.62$ (right) 
selected based on the color--color diagrams (Figure \ref{figure3}).
The solid curves correspond to $\pm 3\sigma$ photometric errors and the
horizontal red lines correspond to $HK - {\rm NB} = 0.4$ [mag], which are
used to select the NB emitters.
The vertical dashed lines correspond to the 5$\sigma$ limiting magnitudes
of the NB images.}
\label{figure2}
\end{figure*}

\begin{figure*}[ht]
\centering\includegraphics[width=0.3\textwidth, angle=270]{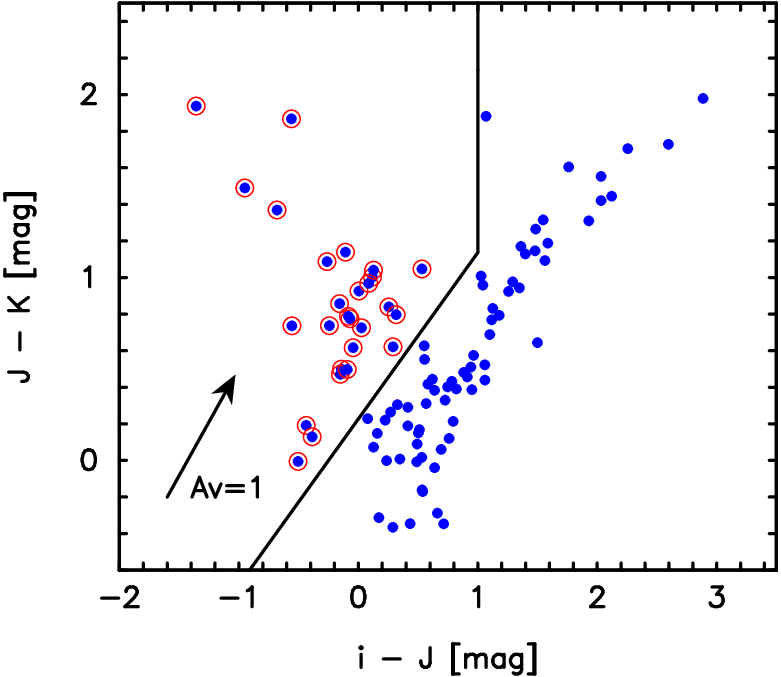}
\hspace{10mm}
\centering\includegraphics[width=0.3\textwidth, angle=270]{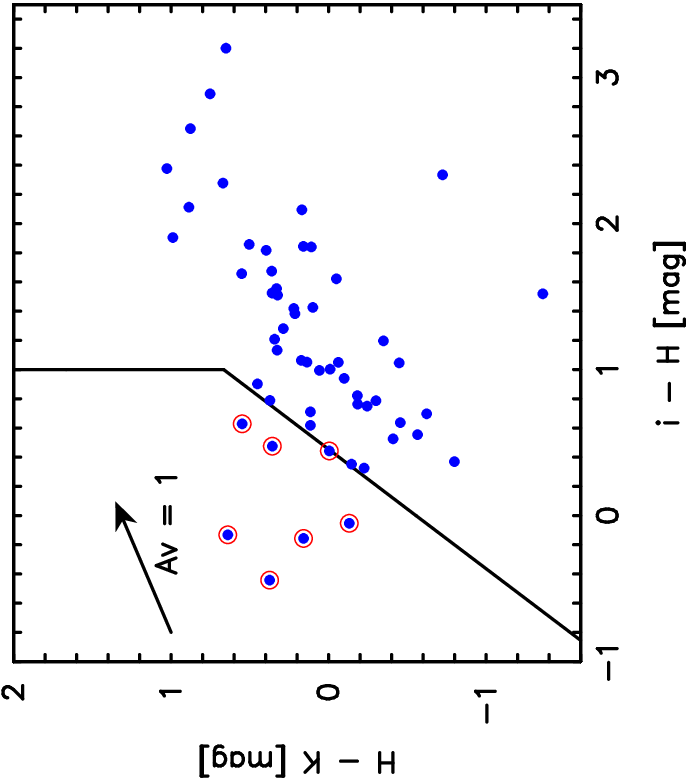}
\caption{The color--color diagrams for NB209 (left) and NB2315 (right). 
Blue filled circles represent the NB emitters selected based on
the color--magnitude diagrams (Figure \ref{figure2}),
and red circles show the \oiii\ emitters at $z=3.17$ (left) and $z=3.62$ (right)
which satisfy the criteria shown by the dividing solid lines.
The arrow represents the reddening vector of $A_{\rm v}=1$ mag.}
\label{figure3}
\end{figure*}

First of all, we extract sources from the NB images taken with MOIRCS and 
the broad-band (BB; {\it H}- and {\it K}- bands) images taken with WFCAM,
using the public software {\tt SExtractor} \citep{bertin96}.
The pixel scales and PSF sizes of the BB images are matched to those of the NB images.
We perform aperture photometries on the NB and BB images
with a $1.6''$ diameter aperture.  
Source extraction and photometries are carried out with the double image mode of {\tt SExtractor}.
Those aperture photometry data are used to select line emitters based on
the color--magnitude diagrams (Figure \ref{figure2}) and to measure NB fluxes.
On the other hand, the template-fitting photometry data from the public catalog 
(see \citealt{galametz13} for more details) are used for the color--color
selections (Figure \ref{figure3}), the spectral energy distribution (SED)
fitting, and the SFR measurements from UV luminosities (in $R_c$-band).

The objects that have large excesses in NB fluxes as compared to BB fluxes
are selected as the NB emitters.
Since the effective wavelengths of the NB filters and the $K$-band are
slightly different, we estimate BB fluxes at the exact effective wavelengths
of the NB209/NB2315 filters by interpolating fluxes between
$H$- and $K$-bands as follows \citep{tadaki13};

\begin{equation}
HK(\lambda = 2.09\ \mu {\rm m}) = 0.8\ K + 0.2\ H - 0.015, 
\end{equation}
\begin{equation}
HK(\lambda = 2.315\ \mu {\rm m}) = 1.2\ K - 0.2\ H + 0.011.
\end{equation}

\noindent
We select NB excess sources using $HK - {\rm NB209}$ or $HK - {\rm NB2315}$ color--magnitude diagrams (Figure \ref{figure2}).
We use a parameter $\Sigma$, which determines the significance of 
a NB excess relative to an 1$\sigma$ photometric error \citep{bunker95}.
The relation between $\Sigma$ and the color of $m_{\rm BB} - m_{\rm NB}$ is obtained from 

\small{ 
\begin{equation}
m_{\rm BB} - m_{\rm NB} = -2.5 \ {\rm log_{10}} \left[ 1 - \frac{\Sigma \sqrt{\sigma^2 _{\rm BB} + \sigma^2 _{\rm NB}}}{f_{\rm NB}} \right],
\end{equation}
}

\normalsize
\noindent
where $f_{\rm NB}$ is a NB flux density, $\sigma^2 _{\rm BB}$ and $\sigma^2 _{\rm NB}$
are the sky noises in the BB and NB images measured within an aperture, respectively \citep{tadaki13}.
We set the criterion of $\Sigma=3$ to sample secure emitters, and it is
represented as the solid curve in Figure \ref{figure2}.
We also set the criteria that the NB magnitude excess with respect to
the $HK$ magnitude, $HK - {\rm NB}$, is larger than 0.4 mag,
and that the NB magnitude is brighter than the 5$\sigma$ limiting
magnitude (the horizontal red line and the vertical dashed line in
Figure \ref{figure2}, respectively).
With these selection criteria, we obtain 101 and 58 NB excess sources
in NB209 and NB2315, respectively.

The NB209(NB2315) filter captures different emission lines from galaxies at different redshifts, 
such as \ha\ emission lines at $z=2.19 (2.53)$ and \oiii\ emission lines at $z=3.17 (3.62)$. 
In order to construct a sample of \oiii\ emitters at $z>3$, we need to 
separate \oiii\ emitters from \ha\ emitters at $z<3$.
The conspicuous spectral features such as the Balmer and/or $4000$\AA\ breaks
can be used for this purpose.
Since the break feature is redshifted to a different wavelength in the observed 
frame according to the redshift of a galaxy, if we choose an appropriate
combination of passbands that can neatly straddle the spectral break feature,
we are able to disentangle those various possible solutions for different
line emitters at different redshifts.
\citet{tadaki13} use $i'-J$ versus $J-K$ diagram for NB209 emitters
and $i'-H$ versus $H-K$ diagram for NB2315 emitters. 
They have shown that \oiii\ emitters at $z>3$ and \ha\ emitters at $z<3$ can be well 
separated by the dividing lines as shown in Figure \ref{figure3}.
We follow their selection method 
and here the contribution from the emission line is subtracted from the $K$-band magnitude.
Our NB emitter sample may also contain some \pa\ emitters at low redshifts
($z=$0.1--0.2).
However, since the \pa\ emitters should appear in the same regions
as the \ha\ emitters on these diagrams, they should not be major
contaminants for our \oiii\ emitters.
We have thus finally obtained strong candidates for \oiii\ emitters;
27 at $z=3.17$ and 7 at $z=3.62$.

\begin{figure}[h]
\centering\includegraphics[width=0.35\textwidth, angle=270]{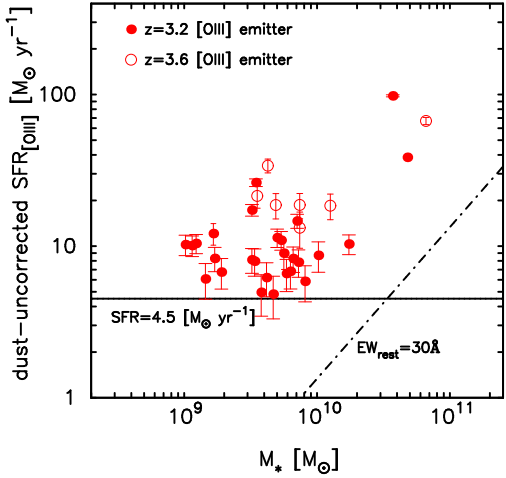}
\caption{The stellar mass versus SFR diagram for the \oiii\ emitters
{\it before dust correction is applied} to discuss any possible selection bias.
Filled and open circles represent the \oiii\ emitters at $z=3.2$ (NB209) and $z=3.6$ (NB2315), respectively. 
SFRs are derived from \oiii\ line luminosities without dust correction here on purpose.
The horizontal solid line and the dot-dashed line correspond to
our selection criteria of the NB emitters, namely, $\Sigma=3$ and $HK-NB=0.4$,
respectively. These numbers correspond to our detection
limits of ${\rm SFR_{[OIII]}} = 4.5\ {\rm [M_\odot yr^{-1}]}$ and
${\rm EW_{rest}}=30\ {\rm [\AA]}$, respectively.
}
\label{figure4}
\end{figure}

Figure \ref{figure4} shows the relation between stellar masses and dust-extinction-uncorrected
SFRs measured from \oiii\ line luminosities for the \oiii\ emitters
in order to verify our sample selection.
We will explain our method of measuring stellar masses and SFRs in the following sections.
Our criteria of determining NB flux excesses, namely $\Sigma > 3$ and $HK-{\rm NB}209 > 0.4$\ [mag]
as shown in Figure \ref{figure2}, correspond to the limits of ${\rm SFR} > 4.5\ {\rm [M_{\odot} yr^{-1}]}$ 
and ${\rm EW_{rest}} > 30\ {\rm [\AA]}$, respectively.
We draw a line corresponding to the EW cut on the $M_*$--SFR diagram 
by establishing a relation between stellar mass and SFR  
along the threshold of $HK-{\rm NB}209=0.4$.
As proxies of the $H-K$ and $J-K$ colors along this boundary line, 
we use the averaged colors of the 3 objects which are located nearest to the boundary of $HK-{\rm NB}209=0.4$.
Then, for a given NB magnitude, we assign a $K$-band magnitude using $HK-{\rm NB}209=0.4$ and the above $H-K$ color.
A stellar mass is then estimated from the $K$-band magnitude and the above $J-K$ color.
$J-K$ color is used to correct for the mass-to-light ratio based on the stellar population synthesis model of \citet{kodama98, kodama99}.
Also a dust-extinction-uncorrected ${\rm SFR_{[OIII]}}$ is calculated from each NB magnitude.
Figure \ref{figure4} shows that 
our EW cut is located well below the actual observed data points, and our selected galaxy
sample is not biased to any particular galaxies on the main sequence diagram.

\section{Physical properties of \oiii\ emitters}

\subsection{AGN Contribution}
Not only hot young massive stars in star-forming regions
but also AGN activities can contribute to the \oiii\ line intensity.
Therefore it is important for us to verify the presence of AGN candidates
in our sample.

\begin{figure}[htp]
\begin{center}
\includegraphics[width=0.32\textwidth, angle=270]{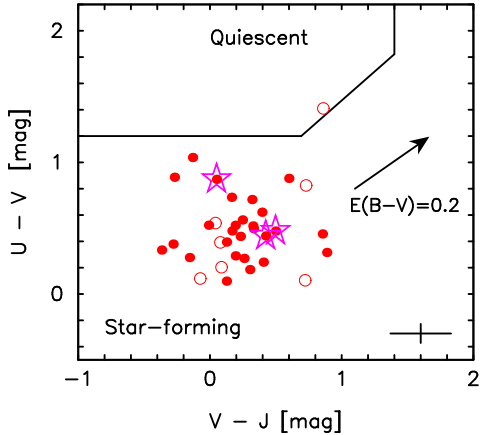}
\caption{The rest-frame $UVJ$ diagram for our \oiii\ emitters
 (filled circles for $z=3.2$ and open circles for $z=3.6$). 
Star marks indicate MIPS detected sources.
The arrow shows a reddening vector of $E(B-V)=0.2$.
The typical errors of the $U-V$ and $V-J$ colors are shown at the bottom-right corner of the figure.
We make 1000 pseudo photometric catalogs for each object 
by randomly generating fluxes in each band from a Gaussian distribution 
with the standard deviation determined by each flux error.
These generated SEDs are fitted by {\tt EAZY} and 
the dispersions of the two colors are estimated for each object.
We show those dispersions as the typical error of each color.
}
\label{figure5}
\end{center}
\end{figure}

We use a photometric redshift code {\tt EAZY} \citep{brammer08} 
to obtain the rest-frame $U$-, $V$-, and $J$-band magnitudes for our sample.
Note that derived photometric redshifts are mostly  
consistent with the expected redshifts, $z=3.2$ and 3.6.
The rest-frame $U - V$ and $V - J$ colors allow us to distinguish 
between two galaxy populations, namely, old quiescent galaxies
and young, dusty star-forming galaxies, by capturing the Balmer/4000\AA\ 
breaks between $U$- and $V$-bands
(e.g. \citealt{wuyts07, williams09, whitaker11}).
Figure \ref{figure5} shows the rest-frame $UVJ$ color--color diagram
for our \oiii\ emitters.
We find that one object is marginally classified as a quiescent galaxy, 
indicating that its \oiii\ emission is likely to be dominated by the AGN
activity rather than the star formation.
However, we cannot discriminate between the contribution from AGNs 
and that from star-forming regions for all the other emitters 
classified as star-forming galaxies.

For further investigation, we inspect the X-ray image by the $XMM-Newton$ \citep{ueda08}.
None of our \oiii\ emitters is detected in X-ray and thus our sample
does not seem to contain any bright unobscured AGNs.
We also look into the $Spitzer$/MIPS 24$\mu$m catalog
and find that three objects are detected with MIPS. 
A fraction of them might be obscured AGNs with warm dust components 
which emit strong IR emissions.



Spectroscopic observations are necessary to confirm 
the presence of AGNs and we do not exclude these objects in the following
analyses.

\subsection{SED fitting}
We perform the SED fitting for our \oiii\ emitters using a public code {\tt FAST} \citep{kriek09}.
We use 18 bands, $u$, $B$, $V$, $R_c$, $i'$, $z'$, $F606W$, $F814W$, $F125W$, $F160W$, 
$Y$, $ J$, $H$, $K_s$, $3.6 \mu$m, 4.5$\mu$m, 5.8$\mu$m, and 8.0$\mu$m.
For the NB209-selected \oiii\ emitters, emission line fluxes are subtracted
from the $K_s$-band fluxes before the SED fitting is performed,
while no correction is required for the NB2315-selected ones as the NB2315
has little overlap with the $K_s$-band in wavelength.
The redshifts of the NB209- and NB2315-selected \oiii\ emitters are fixed
to $z=3.17$ and $z=3.62$, respectively, for the SED fitting.
We use the stellar population synthesis model of \citet{bc03},
the Salpeter IMF \citep{salpeter55}, and the dust attenuation law of \citet{calzetti00}.
We assume the exponentially declining star formation history in the form of
${\rm SFR} \sim {\rm exp}(-t/\tau)$ with ${\rm log(\tau/yr)}$ = 7.0--10.0
in steps of 0.5, and the solar metallicity. 
The output physical quantities from the {\tt FAST} code are star formation
timescale $\tau$, age, dust extinction $A_{\rm V}$, 
stellar mass, SFR, specific SFR and age/$\tau$ ratio.

The stellar masses and dust extinctions ($A_{\rm V}$) used in the
following analyses are all estimated by the SED fitting.


\subsection{Star Formation Rates}
We estimate SFRs of the \oiii\ emitters with the two different indicators,
namely, the UV continuum luminosities (tracing hot young stars)
and the \oiii\ emission line intensities (tracing star-forming HII regions).
In the former case, we adopt the following equation from \citet{madau98}; 

\begin{eqnarray}\nonumber
{\rm SFR \ (M_\odot yr^{-1})} &=& \frac{4\pi D_L^2  f_{\nu}}{(1+z) 8 \times 10^{27} \ ({\rm erg \ cm^{-2} s^{-1} Hz^{-1}})} \\
&=& \frac{L({1600} {\rm \AA})}{8 \times 10^{27} \ ({\rm erg \ s^{-1} Hz^{-1}})}, 
\label{eq4}
\end{eqnarray}

\noindent
where $D_L$ is the luminosity distance and $f_{\nu}$ is the flux density derived from the $R_c$-band magnitude ($\lambda_{\rm c} = 6498.1$\AA).
The dust extinction at $1600$\AA\ is estimated from the SED-based
${A_{\rm V}}$ value and the extinction curve for starburst galaxies of \citet{calzetti00}; 

\begin{equation}
E(B-V)_{\rm stellar} = A'_{\rm V} / R'_{\rm V},
\end{equation}

\begin{equation}
A'(\lambda) = k'(\lambda) E(B-V)_{\rm stellar}, 
\end{equation}

\noindent
and 

\begin{eqnarray}\nonumber
k'(\lambda) =& 2.659 (-2.156 + 1.509/\lambda - 0.198/\lambda ^2 + 0.011/\lambda ^3) \\ &+ R'_{\rm V}, 
  \ \ \ {\rm at} \  0.12 \mu{\rm m} \le \lambda \le 0.63 \mu{\rm m}.
\end{eqnarray}

\noindent
$E(B-V)_{\rm stellar}$ indicates the amount of reddening in the stellar
continuum, and $R'_{\rm v}$ is 4.05 for starburst galaxies \citep{calzetti00}.
The intrinsic flux density $f_i (\lambda)$ is then obtained as

\begin{equation}
f_i(\lambda) = f_o (\lambda) 10^{0.4A'(\lambda)}, 
\end{equation}

\noindent
and the dust-extinction-corrected SFRs (${\rm SFR_{UV}}$) are derived using Eq(\ref{eq4}).

In \citet{maschietto08}, they derive a SFR from an \oiii\ emission line
strength by assuming the \oiii/\ha\ ratio of $\sim 2.4$ which is the
maximum value for local star-forming galaxies \citep{moustakas06}.
Considering the fact that high-$z$ star-forming galaxies show very high
\oiii/\hb\ ratio due to the high excitation states 
(e.g. \citealt{masters14, holden14, steidel14, shimakawa14}), 
this assumption seems to be reasonable for our sample, 
although this ratio has a large dispersion among individual galaxies
\citep{moustakas06}. 
We adopt this maximum ratio to the relation between the SFR and \ha\ luminosity
of \citet{kennicutt98};

\begin{equation}
{\rm SFR_{H\alpha} \ (M_\odot yr^{-1})} = 7.9 \times 10^{-42}\frac{L_{\rm H\alpha}}{{\rm erg\ s^{-1}}}. 
\end{equation}

\noindent
The lower limit of ${\rm SFR_{[OIII]}}$ is thus obtained by;

\begin{equation}
{\rm SFR_{[OIII]} \ (M_\odot yr^{-1})} > 0.33 \times 10^{-41} \frac{L_{\rm [OIII]}}{{\rm erg\ s^{-1}}}.
\end{equation}

\noindent
The luminosity of the \oiii\ emission line, $L_{\rm [OIII]}$, is obtained by measuring the \oiii\ line flux from the NB and BB flux densities.
The NB and BB flux densities are defined as 

\begin{equation}
f_{\rm NB} = f_{c} + F_{\rm line}/\Delta_{\rm NB},
\end{equation}

\begin{equation}
f_{\rm BB} = f_c + F_{\rm line}/\Delta_{\rm BB}, 
\end{equation}

\noindent
where $f_c$ is a continuum flux density, $F_{\rm line}$ is a line flux intensity,
and $\Delta_{\rm NB}$ and $\Delta_{\rm BB}$ are FWHMs of the NB and BB
filters \citep{tadaki13}.
The continuum flux density, the line flux intensity, and the equivalent width (EW) 
in the rest-frame are given by the following equations, respectively;

\begin{equation}
f_c = \frac{f_{\rm BB} - f_{\rm NB} (\Delta_{\rm NB}/\Delta_{\rm BB})}{1 - \Delta_{\rm NB}/\Delta_{\rm BB}},
\end{equation}

\begin{equation}
F_{\rm line} = \Delta_{\rm NB} \frac{f_{\rm NB} - f_{\rm BB}}{1 - \Delta_{\rm NB}/\Delta_{\rm BB}}, 
\end{equation}

\begin{equation}
{\rm EW_{rest}} = \frac{F_{\rm line}}{f_c} (1+z)^{-1}.
\end{equation}

\noindent
The line flux $F_{\rm line}$ is converted to the \oiii\ luminosity with $ L_{\rm [OIII]} = 4 \pi D^2_L F_{\rm line}$.
The dust extinction at 5007\AA\ is estimated in the same manner as used for the dust extinction at 1600\AA\ based on the SED fitting.
We assume that there is no extra extinction for the nebula emissions compared to the stellar extinction, 
i.e. $E(B-V)_{\rm stellar} = E(B-V)_{\rm nebula}$.

\begin{figure}[tbp]
\begin{center}
\includegraphics[width=0.34\textwidth, angle=270]{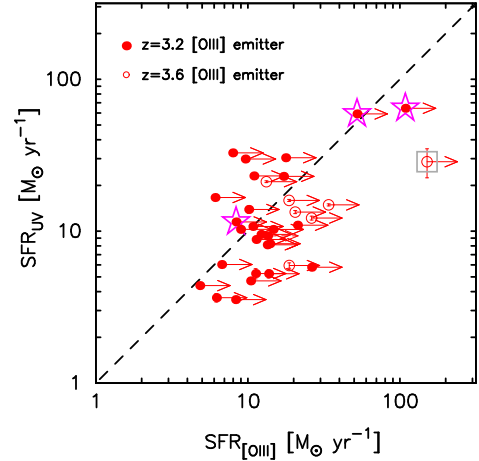}
\caption{The comparison between the dust extinction corrected ${\rm SFR_{[OIII]}}$ and ${\rm SFR_{UV}}$
 (filled circles for $z=3.2$ and open circles for $z=3.6$).
 Arrows represent that ${\rm SFR_{[OIII]}}$ is shown as the lower limit.
Star marks show MIPS detected objects, and 
a square indicates an object classified as a quiescent galaxy based on
the $UVJ$ diagram (Figure \ref{figure5}).
The dashed line represents the case where two measurements are identical.}
\label{figure6}
\end{center}
\end{figure}

In Figure \ref{figure6}, we compare SFRs derived from the two different indicators.
The ratios of ${\rm SFR_{[OIII]}}/{\rm SFR_{UV}}$ range from 0.25 to 3 for most of the objects.
On the other hand, the object classified as a quiescent galaxy on the {\it UVJ} diagram (Figure \ref{figure5}) shows 
a slightly higher ratio of ${\rm SFR_{[OIII]}}/{\rm SFR_{UV}} \sim 5.3$.
It suggests that this object has an extra contribution from an AGN to its \oiii\ emission as expected in Section 3.1.

\subsection{$M_*$--SFR Relation}

\begin{figure}[tbp]
\begin{center}
\includegraphics[width=0.45\textwidth, angle=270]{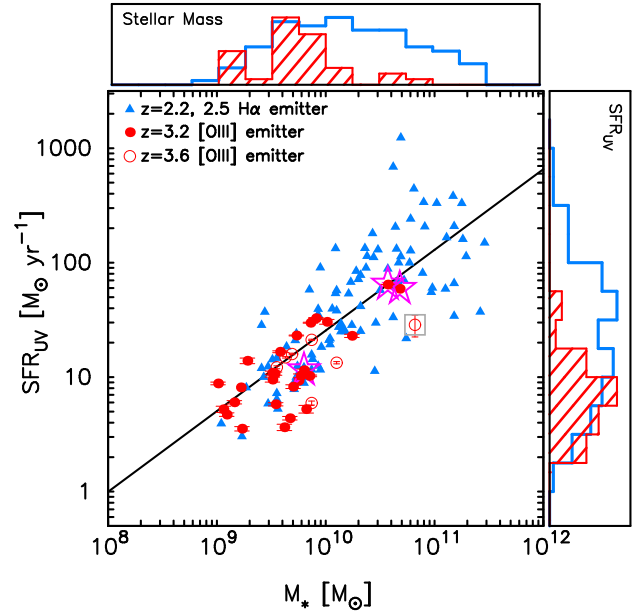}
\caption{The stellar mass and SFR relation (main sequence of star-forming galaxies)
for the \oiii\ emitters (filled circles for $z=3.2$ and open circles for $z=3.6$) 
and the \ha\ emitters at $z$=2.2 and 2.5 from \citet{tadaki13}
(filled triangles).
SFRs are derived from UV luminosities and corrected for dust extinction.
Errors in SFRs for the \oiii\ emitters are estimated from 1$\sigma$ photometric errors of the $R_c$-band magnitudes.
The solid line represents the best-fit line to the \ha\ emitters; 
${\rm SFR_{UV}} = 129 \ M_{11}^{0.705}$ ($M_{11} = M_*/10^{11}{\rm M_\odot}$). 
Top and right-side histograms show the stellar mass and SFR distributions 
of the \oiii\ emitters at $z$=3.2 and 3.6 (red hatched histograms) and the \ha\ emitters at $z$=2.2 and 2.5 (blue open histograms). 
}
\label{figure7}
\end{center}
\end{figure}

We investigate the relation between stellar masses and SFRs 
(the ``main sequence'' of star-forming galaxies) for the \oiii\ emitters at $z>3$.
The dust-extinction-corrected SFRs (${\rm SFR_{UV}}$) of most of the \oiii\ emitters
at SXDF range from a few to $\sim$ 30 ${\rm M_\odot yr^{-1}}$.
Figure \ref{figure7} shows a $M_*$--SFR relation for the \oiii\ emitters
at $z$=3.2 and 3.6, together with the \ha\ emitters at $z$=2.2 and 2.5
in the same field \citep{tadaki13}.
We re-estimate stellar masses and dust-extinction-corrected SFRs (${\rm SFR_{UV}}$)
of the \ha\ emitters in the same manner as in this study by using the
{\tt FAST} code (Section\ 3.2).

We find that stellar masses and SFRs of the \oiii\ emitters show a clear
correlation as seen in other studies of star-forming galaxies across a wide redshift range 
(e.g. \citealt{elbaz07, daddi07, whitaker12, koyama13, kashino13, tasca14}).
The normalization of the $M_*$--SFR relation for the \oiii\ emitters at $z$=3.2 and 3.6 
looks almost identical to that of the \ha\ emitters at $z$=2.2 and 2.5.
We confirm that the best-fit line to the \oiii\ emitters is consistent with 
the fit to the \ha\ emitters within 1$\sigma$ errors in both the slopes and the intercepts. 
Importantly, however, the distributions of galaxies along the main sequence are systematically 
different in the sense that the stellar masses of the \oiii\ emitters at $z\sim3.2$ and 3.6 are much 
(nearly by a factor of 10) lower than those of the \ha\ emitters at $z\sim2.2$ and 2.5.
We can simply interpret the difference in two ways; 
(1) the evolution of galaxies from $z$=3.2, 3.6 to $z$=2.2, 2.5, 
and 
(2) the selection bias between \oiii\ and \ha\ emitters. 
In Section 4.1, we will refer to the option (2), and  
in Section 4.2, we will assume that the difference between stellar mass distributions is only due to 
the evolution of galaxies from $z$=3.2, 3.6 to $z$=2.2, 2.5, and 
discuss how their stellar masses and SFRs grow in this time interval. 

\subsection{Sizes}
Our NB imaging survey areas are covered by the $HST$/CANDELS fields, and
high spatial resolution images in the rest-frame UV-optical wavebands
(ACS and WFC3) are available for the \oiii\ emitters.

We use the structural parameters measured by \citet{vanderwel12}
on the $H_{F160W}$-band selected objects in CANDELS.
We briefly summarize below their methods to obtain the structural parameters.
They perform a S\'ersic model fit to the $H_{F160W}$-band selected objects
using the softwares; {\tt GALAPAGOS} \citep{barden12} and
{\tt GALFIT} \citep{peng10}.
The parameters that are used for the fit are the total magnitude, half-light radius measured along the major axis, 
S\'ersic index, axial ratio, position angle, and central position.
The initial guesses of these parameters are given by {\tt SExtractor}.
The best-fit {\tt GALFIT} parameters for all the objects are publicly available
in the Rainbow Database \citep{galametz13}.
A flag number between 0 and 3 is assigned for each object.  
We reject objects with ${\rm flag}\ge 2$ because the fitting result with a S\'ersic model becomes increasingly unreliable.
They note that resultant structural parameters have systematic and random uncertainties
and that such uncertainties depend on the brightness of objects and become larger for fainter objects.
At $H_{F160W} = 25$, systematic and random uncertainties of the half-light radius ($r_e$ in arcsec)
are $0.04''$ and $0.18''$, respectively, when $r_e$ is less than 0.3$''$,
while they are $-$0.09$''$ and 0.33$''$ when $r_e$ is greater than 0.3$''$.
At $H_{F160W} = 26$, those uncertainties become as large as $0.12''/0.42''$ ($r_e<0.3''$)
and $-0.11''/0.63''$ ($r_e > 0.3''$), respectively.

Using the {\tt GALFIT} parameters from \citet{vanderwel12}, we estimate the
effective radius $r_e$\ [kpc] in the rest-frame $U$-band for the \oiii\ emitters.
We use only the bright objects ($H_{F160W} < 25$) with the flag values of 0 or 1.
Based on the magnitude cut and the flag values, 10 and 9 objects, respectively, 
are excluded from the \oiii\ emitter sample.


\subsection{$M_*$--Size Relation}
\begin{figure}[tbp]
\begin{center}
\includegraphics[width=0.4\textwidth, angle=270]{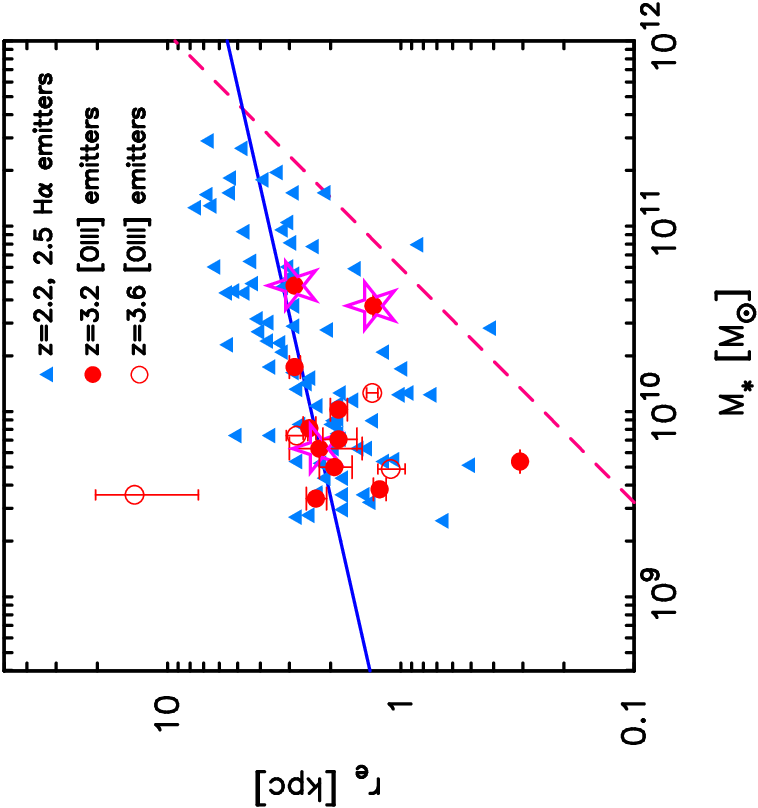}
\caption{The stellar mass and size relation for the \oiii\ emitters (filled circles: $z=3.2$, open circles: $z=3.6$) with $H_{F160W} < 25$ 
and the \ha\ emitters at $z$=2.2 and 2.5 from \citet{tadaki13} (filled triangles) with $J_{F125W} < 25$.
The solid and dashed lines represent the mass-size relations 
of $z \sim 2.75$ late- and early-type galaxies, respectively \citep{vanderwel14}.
Note that their galaxy sizes in \citet{vanderwel14} are estimated at a rest-frame wavelength of 5000\AA, 
slightly longer wavelength than the rest-frame $U$-band where our galaxy sizes are measured.
}
\label{figure8}
\end{center}
\end{figure}

Figure \ref{figure8} shows the relation between stellar masses and sizes for the \oiii\ emitters at $z>3$.
The \ha\ emitters at $z=2.2$ and 2.5 are also shown. 
Their sizes are estimated in the {\it HST} $J_{125W}$-band images 
so that they can be directly compared to those of our \oiii\ emitters
at the same rest-frame wavelength.
The size measurements are also limited to the bright objects ($J_{F125W} < 25$)
with ${\rm flag=0\ or\ 1}$.
In Figure \ref{figure8}, the solid and dashed lines represent the mass--size relations 
of late- and early- type galaxies at $2.5 < z_{\rm phot} < 3$, respectively, 
derived from the 3D-{\it HST}/CANDELS group \citep{vanderwel14}.
We find that the size distribution of the \oiii\ emitters with respect to the stellar mass is similar to that of the \ha\ emitters, 
and that they follow the mass--size relation of late-type galaxies at $z\sim2.75$ from \citet{vanderwel14}.
In \citet{vanderwel14}, galaxy sizes are estimated at a rest-frame wavelength of 5000\AA, 
slightly longer wavelength than the rest-frame $U$-band where our galaxy sizes are measured.
To verify a possible effect due to wavelength mismatch, 
we also apply their same correction method to our sample, 
and confirm that there is no systematic difference between the two measurements at different wavelengths.

While most of the \oiii\ emitters have sizes consistent with the mass--size relation of late-type galaxies at $z\sim2.75$, 
there is a massive \oiii\ emitter for its size ($M_* \sim 3 \times 10^{10} {\rm M_\odot}$ and $r_e \sim 1\ \rm{kpc}$).
Massive and compact star-forming galaxies are expected to evolve to massive and compact quiescent galaxies 
when their star formation is quenched (e.g. \citealt{barro13}; \citealt{tadaki14}).
We confirm the presence of such massive and compact star-forming galaxies at $z=3.2$.

\section{Discussions}
\subsection{Selection Bias}
We note here on a possible selection bias introduced by our use of the \oiii\ emission line
as an indicator of star-forming galaxies. 
When we use the \oiii\ line,  
the galaxy sample tend to be biased towards galaxies with more extreme ISM conditions.
It has been found that high-$z$ star-forming galaxies tend to have much higher excitation 
states (e.g. \citealt{masters14, holden14, steidel14, shimakawa14}).
\citet{shimakawa14} perform the NIR spectroscopic observations of the \ha\ emitters
at $z$=2.2 and 2.5 associated to the two protocluster fields, and have shown that the
\oiii/\ha\ ratios measured from the stacked spectra are $\sim$1.0--3.0 
in the stellar mass range of $10^{9}$--$10^{11} {\rm M_\odot}$.
The extreme ISM condition is expected to be a common feature among high-$z$
star-forming galaxies, and we expect that the \oiii\ emission line is an appropriate
tracer of normal star-forming galaxies at high redshifts.

The \oiii\ emitters may also be biased to less dusty galaxies compared to the \ha\ emitters,
since the \oiii\ emission line (5007\AA) is located at the slightly shorter wavelength than
the \ha\ emission line (6563\AA) and hence more strongly affected by dust extinction.
However, adopting the extinction curve of \citet{calzetti00}, 
the dust extinction at the wavelength of the \oiii\ line is only $\sim 1.3$ times larger 
than that at the wavelength of the \ha\ line.
Considering that high-$z$ star-forming galaxies tend to have high \oiii/\ha\ ratios
as mentioned above, the effect of dust extinction for the \oiii\ line would not introduce 
a strong bias to less dusty galaxies.

Moreover, metallicity of galaxies may also affect the strength of \oiii\ emission. 
Since lower metallicity leads to higher stellar temperature, 
\oiii\ line becomes stronger. 
Given the well known mass-metallicity relation of star-forming galaxies (e.g. \citealt{erb06a}), 
this metallicity effect may result in a possible bias towards lower stellar masses for \oiii\ emitters 
as compared to \ha\ emitters. 

In order to verify those selection biases,  
HiZELS (the High-redshift(Z) Emission Line Survey; \citealt{best10, sobral13, sobral14}) 
offers a very unique sample of dual emitters.
They used a pair NB filters to capture \oiii\ and \ha\ emission lines at the same redshift, 
and constructed the samples of \oiii\ emitters and \ha\ emitters at $z=2.23$.
We will address the selection biases between the two samples based 
on this unique data sets in a forthcoming paper (D. Sobral, private communication). 


\subsection{Galaxy Growth from $z=3.2$ to $z=2.2$}
In Section 3.4, we show that there is no significant change in the location of the main sequence 
of star-forming galaxies between $z=3.2$ (3.6) and $z=2.2$ (2.5), 
but 
 the galaxy distributions on the sequence are different between the two epochs. 
In this section, we assume that the difference in galaxy distributions on the $M_*$--SFR plane 
between the \oiii\ emitters and \ha\ emitters is simply due to the evolution of star-forming galaxies between the two epochs 
and  discuss the stellar mass growth of galaxies from $z=3.2$ to $z=2.2$.  
From our result that the location of the main sequence is unchanged during this time
interval (1Gyr), which is represented by ${\rm SFR} = 129 M_{11}^{0.705}$ as defined
for the \ha\ emitters at $z$=2.2 and 2.5 (Figure \ref{figure7}), 
we can put some constraints on the history of star formation and thus that of the stellar
mass growth. 
In order to stay on the same main sequence, the simplest evolutionary path would be
that the individual star-forming galaxies evolve {\it along} the main sequence.
This assumption should be valid if the galaxies keep forming stars at the rates
above our threshold of the \ha\ NB imaging,
i.e. ${\rm SFR} > 4\ {\rm [M_\odot yr^{-1}]}$ (dust-uncorrected) and ${\rm EW_{rest}(H\alpha + [N_{II}]) > 40\ [\AA]}$.

The stellar mass growth between $z=3.2$ and $z=2.2$ can be approximately tracked
by the following derivative equation;

\begin{equation}
 {\rm d}M_*/{\rm d}t = (1-R) \times {\rm SFR} = (1-R) \times 129\ M_{11}^{0.705}, 
\end{equation}

\noindent
where the return mass fraction $R$ is $\sim0.3$ for the Salpeter IMF.
Using this equation, a galaxy with $M_* = 10^9 {\rm M_{\odot}}$ at $z=3.2$ can  
increase their stellar mass by a factor of 10 to
$\sim 1.1 \times 10^{10} {\rm M_{\odot}}$ by $z=2.2$, while
a galaxy with $M_* = 10^{11} {\rm M_{\odot}}$ can grow in mass by a factor of 2.
Therefore, more than 50\% up to $\sim 90 \%$ of the stellar mass of the star-forming
galaxies at $z=2.2$ can be formed during the 1 Gyr time interval between $z=3.2$ and 2.2.
The majority of the galaxies with $M_* > 10^9 {\rm M_\odot}$ that we see at $z=3.2$ 
would grow to massive galaxies of $M_* > 10^{10} {\rm M_\odot}$ at $z=2.2$, 
if they keep their high star formation activities.

Note also that, in this simple model, 
galaxies {\it climb up} the main sequence, i.e. 
SFR increases a lot from $z=3.2$ to $z=2.2$ as the stellar mass grows.
This indicates that the star formation activities of galaxies at $z>3$ are accelerated 
towards the peak epoch of galaxy formation at $z\sim2$.  
In this respect, $z>3$ is the {\it pre-peak epoch} of galaxy formation. 
In order to achieve such an increasing star formation activity, an increasing rate of gas
infall from outside is required, since otherwise the quick gas consumption would lower SFR
as time progresses.
In order to verify the presence of such continuous gas infall more quantitatively, 
we estimate the gas mass for the \oiii\ emitters from their SFR surface densities 
by assuming the Schmidt-Kennicutt relation \citep{kennicutt98}. 
SFR surface densities ($\Sigma_{\rm SFR} = {\rm SFR}/\pi r_e^2$) are estimated 
by using SFRs derived from UV luminosities in Section 3.3 and the effective radius $r_e$ in Section 3.5.
We then calculate the gas depletion time-scale of $t_{\rm dep} = M_{\rm gas}/{\rm SFR}$.
The depletion time-scale of the \oiii\ emitters is mostly in the range of $0.2-0.4$ Gyr, and shorter than 1 Gyr.
This means that the \oiii\ emitters at $z=3.2$ would consume all the remaining gas and 
terminate the star formation before $z=2.2$ 
if there is no gas supply from the outside of galaxies.

We have to mention that we have assumed the exponentially {\it declining} star formation history (SFH) 
in the form of ${\rm SFR} \sim {\rm exp}(-t/\tau)$ in the SED fitting, 
while we now claim that the SFR {\it increases} with time from $z=3.2$ to $z=2.2$. 
In order to verify the impact of assumed SFHs on the resulting physical quantities in the SED fitting, 
we re-estimate the stellar masses and SFRs of the \oiii\ emitters by assuming the exponentially {\it increasing} SFH.
In the case of the increasing SFH, the estimated stellar masses vary by only factor of $0.9-1.3$ for most of our sample, 
while the SFRs derived with $A_{\rm V}$ values from the SED fitting can increase by a factor of $\sim$ 1.4.
However, such a modest offset would be systematic and would apply to both the \ha\ and \oiii\ emitter samples. 
Therefore it should not change our results significantly.

In reality, some galaxies would stop their star formation and evolve to quiescent
galaxies by $z=2.2$, 
although this quenching process should happen on a relatively short timescale 
so that they do not significantly appear on the lower side of the main sequence
and break its clear sequence.
Also, we have ignored the effect of galaxy-galaxy mergers which can also increase
the stellar mass of galaxies.
Moreover, some galaxies would pop out all of a sudden on the main sequence
with $M_* > 10^{10} {\rm M_{\odot}}$ sometime between $z=3.2$ and $z=2.2$,
which were below $M_* \le 10^{9} {\rm M_{\odot}}$ at $z=3.2$ or somewhere off the main sequence.
Those galaxies should form stars at even higher rates such as in a starburst mode,
and the fraction of stars that are formed between the two epochs can be larger than 90\%.

The presence of those missing galaxies that are not considered in the simple model
above is indicated by the comparison of number densities of the \oiii\ emitters at $z=3.2$
and the \ha\ emitters at $z=2.2$.
The number density of the \oiii\ emitters at $z$=3.2 with $M_* \geq 10^{9} {\rm M_\odot}$ 
is $1.7 \times 10^{-3}\ {\rm [Mpc^{-3}]}$, 
while that of the \ha\ emitters at $z$=2.2 with $M_* \geq 10^{10} {\rm M_\odot}$
is $2.7 \times 10^{-3}\ {\rm [Mpc^{-3}]}$.
Here we have taken into account the mass growth predicted by the above simple model. 
The latter number is $\sim$1.6 times larger. 
It suggests that, some galaxies may actually appear on the main sequence suddenly between $z=3.2$ and $z=2.2$, 
if we consider that there is no selection bias between the \oiii\ emitters and the \ha\ emitters (see Section 4.1).


In any case, it is likely that 
star-forming galaxies grow at an accelerated pace during this time interval, 
assuring that this epoch is critically important for galaxy formation.

We also investigate the size growth of galaxies from $z=3.2$ to $z=2.2$ 
by assuming that the mass-size relation is unchanged between $z=3.2$ and $z=2.2$
as suggested in Section 3.6.
Using the mass--size relation of late-type galaxies at $z\sim2.75$ from \citet{vanderwel14}, 
the effective radius of a galaxy with $M_* = 10^9 {\rm M_{\odot}}$ at $z=3.2$
would grow in size by a factor of $\sim$ 1.5 by $z=2.2$, 
and the size growth ratio does not depend much on the initial stellar mass of galaxies at $z=3.2$.
The size growth is not so strong from $z$=3.2 to $z$=2.2 as compared to the mass
growth that we just discussed above.
Considering the growth of the stellar mass and the size of galaxies from $z=3.2$ to $z=2.2$ together, 
we can also estimate the evolution in stellar mass surface density. 
It is predicted to grow by a factor of 5 for a galaxy with $M_*=10^9 {\rm M_\odot}$ at $z=3.2$.

\section{Summary}
In this study, we construct an \oiii\ emitter sample at $z>3$ in SXDF from the NB imaging
data taken with MOIRCS on the Subaru telescope \citep{tadaki13}.
We identify 27 and 7 \oiii\ emitters at $z$ = 3.2 and 3.6, respectively.
Some objects in our \oiii\ emitter sample might be contributed by AGNs 
based on  the rest-frame {\it UVJ} diagram and the {\it Spitzer}/MIPS detections.
The spectroscopic observation is required to confirm the presence of AGNs and 
we do not exclude these objects in this study.
Using the multi-wavelength data and {\it HST} high resolution images, 
we investigate their basic physical properties, and compared them with those of
the \ha\ emitters at $z$ = 2.2 and 2.5 in the same field.

\begin{itemize}


\item The stellar mass and the dust-extinction-corrected ${\rm SFR_{UV}}$ of the \oiii\ emitters show a clear correlation 
as seen in other previous studies over a wide redshift range.
Comparing our \oiii\ emitters at $z=3.2$ and 3.6 with the \ha\ emitters at $z =2.2$ and 2.5 in the same
field from \citet{tadaki13}, 
the location of the $M_*$--SFR relation of the \oiii\ emitters at $z=3.2$ and 3.6 is 
almost the same as that of the \ha\ emitters at $z=2.2$ and 2.5. 

\item 
Although the location of the relation is almost the same between the \oiii\ and \ha\ emitters, 
the galaxy distributions on the $M_*$--SFR plane are different 
in the sense that  
the \oiii\ emitters at $z=3.2$ and 3.6 tend to have lower stellar masses and SFRs  
as compared to the \ha\ emitters at $z=2.2$ and 2.5. 

\item 
If we assume that the different galaxy distributions on the main sequence are 
due to the evolution of star-forming galaxies from $z=3.2$ to $z=2.2$, 
and that star-forming galaxies simply evolve along the constant star-forming main sequence in this time interval,  
galaxies with $M_* = 10^{9}$--$10^{11} {\rm M_\odot}$ can obtain $\sim$ 90--50\% of their 
stellar masses within just a Gyr from $z=3.2$.
Galaxies {\it climb up} the main sequence, 
and their star formation rates also increase a lot as their stellar masses grow.
Although we consider only the simple model without outflows or mergers,
we infer that galaxy formation activities at $z > 3$ are accelerated towards its peak epoch at $z\sim2$.

\item We investigate the sizes of the \oiii\ emitters measured from the {\it HST} {\it H}-band
images \citep{vanderwel12}.
The size distribution of the \oiii\ emitters at $z=3.2$ and 3.6 
with respect to the stellar mass is similar to that of the \ha\ emitters at $z=2.2$ and 2.5, 
and to that of the late-type galaxies at $z\sim2.75$ from \citet{vanderwel14}. 
When the size of a galaxy grows from $z=3.2$ to $z=2.2$ along the mass--size relation
at $z\sim2.75$ from \citet{vanderwel14}, 
the effective radius would become 1.5 times larger at $z=2.2$, and the size growth
ratio does not depend much on the stellar mass of galaxies at $z$=3.2.
We conclude that the size evolution is not strong from $z=3.2$ to $z=2.2$.
\end{itemize}

\acknowledgments
We thank the anonymous referee for his/her careful reading and comments which improved the clarity of this paper. 
This paper is based on data collected at the Subaru Telescope, 
which is operated by the National Astronomical Observatory of Japan.
We thank the Subaru telescope staff for their great help in the observations.
This work has made use of the Rainbow Cosmological Surveys Database, 
which is operated by the Universidad Complutense de Madrid (UCM), 
partnered with the University of California Observatories at Santa Cruz (UCO/Lick,UCSC).



{\it Facilities:} \facility{Subaru}.

\clearpage




\begin{thebibliography}{}
\bibitem[Ashby et al.(2013)]{ashby13} Ashby, M. L. N., Willner, S. P., Fazio, G. G., et al. 2013, \apj, 769, 80
\bibitem[Barden et al.(2012)]{barden12} Barden, M., H\"au\ss ler, B., Peng, C. Y., McIntosh, D. H., \& Guo, Y. 2012, \mnras, 422, 449
\bibitem[Barro et al.(2013)]{barro13} Barro, G., Faber, S. M., P\'erez-Gonz\'alez, P. G., et al. 2013, \apj, 765, 104
\bibitem[Bertin \& Arnouts(1996)]{bertin96} Bertin, E., \& Arnouts, S. 1996, \aaps, 117, 393 
\bibitem[Best et al.(2010)]{best10} Best, P., et al. 2010, UKIRT 30 Proceedings, preprint (arXiv: 1003.5183)
\bibitem[Bouwens et al.(2012)]{bouwens12} Bouwens, R. J., Illingworth, G. D., Oesch, P. A., et al. 2012, \apj, 754, 83
\bibitem[Brammer et al.(2008)]{brammer08} Brammer, G., van Dokkum, P. G., \& Coppi, P. 2008, \apj, 686, 1503
\bibitem[Bruzual \& Charlot(2003)]{bc03} Bruzual, G., \& Charlot, S. 2003, \mnras, 344, 1000 
\bibitem[Bunker et al.(1995)]{bunker95} Bunker, A. J., Warren, S. J., Hewett, P. C., \& Clements, D. L. 1995, \mnras, 273, 513
\bibitem[Calzetti et al.(2000)]{calzetti00} Calzetti, D., Armus, L., Bohlin, R. C., et al. 2000, \apj, 533, 682
\bibitem[Daddi et al.(2005)]{daddi05} Daddi, E., Renzini, A., Pirzkal, N., et al. 2005, \apj, 626, 680
\bibitem[Daddi et al.(2007)]{daddi07} Daddi, E., Dickinson, M., Morrison, G., et al. 2007, \apj, 670, 156
\bibitem[Damjanov et al.(2009)]{damjanov09} Damjanov, I., McCarthy, P. J., Abraham, R. G., et al. 2009, \apj, 695, 101
\bibitem[Elbaz et al.(2007)]{elbaz07} Elbaz, D., Daddi, E., Le Borgne, D., et al. 2007, \aap, 468, 33
\bibitem[Erb et al.(2006a)]{erb06a} Erb, D. K., Shapley, A. E., Pettini, M., et al. 2006a, \apj, 644, 813
\bibitem[Erb et al.(2006b)]{erb06} Erb, D. K., Steidel, C. C., Shapley, A. E. et al. 2006b, \apj, 647, 128
\bibitem[Fan et al.(2004)]{fan04} Fan, X., Hennawi, J. F., Richards, G. T., et al. 2004, \aj, 128, 515
\bibitem[Fontana et al.(2014)]{fontana14} Fontana, A., Dunlop, J. S., Paris, D., et al. 2014, \aap, 570, A11
\bibitem[F\"orster Schreiber et al.(2009)]{forsterschreiber09} F\"oster Schreiber, N. M., Genzel, R., Bouch\'e, N., et al. 2009, \apj, 706, 1364 
\bibitem[Furusawa et al.(2008)]{furusawa08} Furusawa, H., Kosugi, G., Akiyama, M., et al. 2008, \apjs, 176, 1
\bibitem[Galametz et al.(2013)]{galametz13} Galametz, A., Grazian, A., Fontana, A., et al. 2013, \apjs, 206,10
\bibitem[Gonz\'alez et al.(2010)]{gonzalez10} Gonz\'alez, V., Labb\'e, I., Bouwens, R. J., et al. 2010, \apj, 713, 115
\bibitem[Grogin et al.(2011)]{grogin11} Grogin, N. A., Kocevski, D. D., Faber, S. M., et al. 2011, \apjs, 197, 35
\bibitem[Holden et al.(2014)]{holden14} Holden, B. P., Oesch, P. A., Gonzalez, V. G., et al. 2014, arXiv:1401.5490
\bibitem[Hopkins \& Beacom(2006)]{hopkins06} Hopkins, A. M., \& Beacom, J. F. 2006, \apj, 651, 142
\bibitem[Kashino et al.(2013)]{kashino13} Kashino, D., Silverman, J. D., Rodighiero, G., et al. 2013, \apjl, 777, L8
\bibitem[Kennicutt(1998a)]{kennicutt98a} Kennicutt, R. C., Jr. 1998a, \apj, 498, 541
\bibitem[Kennicutt(1998b)]{kennicutt98} Kennicutt, R. C., Jr. 1998b, \araa, 36, 189
\bibitem[Kodama et al.(1998)]{kodama98} Kodama, T., Arimoto, N., Barger, A. J., \& Arag\'on-Salamanca 1998, \aap, 334, 99
\bibitem[Kodama et al.(1999)]{kodama99} Kodama, T., Bell, E. F., \& Bower, R. G. 1999, \mnras, 302, 152
\bibitem[Kodama et al.(2013)]{kodama13} Kodama, T., Tadaki, K.-i., Hayashi, M., et al. 2013, in IAU Symp. 295, 
The Intriguing Life of Massive Galaxies, ed. D. Thomas, A. Pasquali, \& I. Ferreras (Cambridge: Cambridge Univ. Press), 74
\bibitem[Koekemoer et al.(2011)]{koekemoer11} Koekemoer, A. M., Faber, S. M., Ferguson, H. C., et al. 2011, \apjs, 197, 36
\bibitem[Koyama et al.(2013)]{koyama13} Koyama, Y. Smail, I., Kurk, J., et al. 2013, \mnras, 434, 423
\bibitem[Kriek et al.(2009a)]{kriek09} Kriek, M., van Dokkum, P. G., Labb\'e I., et al. 2009a, \apj, 700, 221
\bibitem[Kriek et al.(2009b)]{kriek09b} Kriek, M., van Dokkum, P. G., Franx, M., Illingworth, G. D., \& Magee, D. K. 2009b, \apjl, 705, L71
\bibitem[Lawrence et al.(2007)]{lawrence07} Lawrence, A., Warren, S. J., Almaini, O., et al. 2007, \mnras, 379, 1599
\bibitem[Madau et al.(1998)]{madau98} Madau, P., Pozzetti, L., \& Dickinson, M. 1998, \apj, 498, 106
\bibitem[Maschietto et al.(2008)]{maschietto08} Maschietto, F., Hatch. N. A., Venemans, B. P., et al. 2008, \mnras, 389, 1223
\bibitem[Masters et al.(2014)]{masters14} Masters, D., McCarthy, P., Siana, B., et al. 2014, \apj, 785, 15
\bibitem[Moustakas et al.(2006)]{moustakas06} Moustakas, J, Kennicutt, R. C., Jr., \& Terminate, C. A. 2006, \apj, 642, 775 
\bibitem[Meurer et al.(1999)]{meurer99} Meurer, G. R., Heckman, T. M., \& Calzetti, D. 1999, \apj, 521, 64
\bibitem[Nakajima \& Ouchi(2014)]{nakajima14} Nakajima, K., Ouchi, M. 2014, \mnras, 442, 900
\bibitem[Noeske et al.(2007)]{noeske07} Noeske, K. G., Weiner, B. J., Faber, S. M., et al. 2007, \apj, 660, L43
\bibitem[Oke \& Gunn(1983)]{oke83} Oke, J. B., \& Gunn, J. E. 1983, \apj, 266, 713
\bibitem[Peng et al.(2010)]{peng10} Peng, C. Y., Ho, L. C., Impey, C. D., \& Rix, H.-W. 2010, \aj, 139, 2097
\bibitem[Reddy et al.(2012)]{reddy12} Reddy, N. A., Pettini, M., Steidel, C. C., et al. 2012, \apj, 754, 25
\bibitem[Salpeter(1955)]{salpeter55} Salpeter, E. E. 1955, \apj, 121, 161
\bibitem[Shen et al.(2003)]{shen03} Shen, S., Mo, H. J., White, S. D. M., et al. 2003, \mnras, 343, 978
\bibitem[Shimakawa et al.(2014)]{shimakawa14} Shimakawa, R., Kodama, T., Tadaki, K.-i., et al. 2014, arXiv:1406.5219
\bibitem[Sobral et al.(2013)]{sobral13} Sobral, D., Smail, I., Best, P. N., et al. 2013, \mnras, 428, 1128
\bibitem[Sobral et al.(2014)]{sobral14} Sobral, D., Best, P. N., Smail, I., et al. 2014, \mnras, 437, 3516
\bibitem[Stark et al.(2009)]{stark09} Stark, D. P., Ellis, R. S., Bunker, A., et al. 2009, \apj, 697, 1493
\bibitem[Stark et al.(2013)]{stark13} Stark, D. P., Schenker, M. A., Ellis, R. S., et al. 2013, \apj, 763, 129
\bibitem[Steidel et al.(2014)]{steidel14} Steidel, C. C., Rudie, G. C., Strom, A. L., et al. 2014, arXiv:1405.5473
\bibitem[Suzuki et al.(2008)]{suzuki08} Suzuki, R., Tokoku, C., Ichikawa, T., et al. 2008, \pasj, 60, 1347 
\bibitem[Tadaki et al.(2011)]{tadaki11} Tadaki, K.-i., Kodama, T., Koyama, Y., et al. 2011, \pasj, 63, 437
\bibitem[Tadaki et al.(2013)]{tadaki13} Tadaki, K.-i., Kodama, T., Tanaka, I., et al. 2013, \apj, 778, 114
\bibitem[Tadaki et al.(2014)]{tadaki14} Tadaki, K.-i., Kodama, T., Tanaka, I., et al. 2014, \apj, 780, 77
\bibitem[Tasca et al.(2014)]{tasca14} Tasca, L. A. M., Le F\'evre, O., Hathi, N. P., et al. 2014, arXiv:1411.5687
\bibitem[Tanaka et al.(2011)]{tanaka11} Tanaka, I., Breuck, C. D., Kurk, J. D., et al. 2011, \pasj, 63, 415
\bibitem[Troncoso et al.(2014)]{troncoso14} Troncoso, P., Maiolino, R., Sommariva, V., et al. 2014, \aap, 563, 58 
\bibitem[Ueda et al.(2008)]{ueda08} Ueda, Y., Watson, M. G., Stewart, I. M., et al. 2008, \apjs, 179, 124
\bibitem[van Dokkum et al.(2008)]{vandokkum08} van Dokkum, P. G., Franx, M., Kriek, M., et al. 2008, \apj, 677, L5
\bibitem[van Dokkum et al.(2010)]{vandokkum10} van Dokkum, P. G., Whitaker, K. E., Brammer, G., et al. 2010, \apj, 709, 1018
\bibitem[van der Wel et al.(2012)]{vanderwel12} van der Wel, A., Bell, E. F., H\"aussler, B., et al. 2012, \apjs, 203, 24
\bibitem[van der Wel et al.(2014)]{vanderwel14} van der Wel, A., Franx, M., van Dokkum, P. G., et al. 2014, \apj, 788, 28
\bibitem[Whitaker et al.(2011)]{whitaker11} Whitaker, K. E., Labb\'e, I., van Dokkum, P. G., et al. 2011, \apj, 735, 86
\bibitem[Whitaker et al.(2012)]{whitaker12} Whitaker, K. E., van Dokkum, P. G., Brammer, G., \& Franx, M. 2012, \apjl, 754, L29 
\bibitem[Williams et al.(2009)]{williams09} Williams, R. J., Quadri, R. F., Franx, M., et al. 2009, \apj, 691, 1879
\bibitem[Wuyts et al.(2007)]{wuyts07} Wuyts, S., Labb\'e, I., Franx, M., et al. 2007, \apj, 655, 51
\bibitem[Wuyts et al.(2011)]{wuyts11} Wuyts. S., F\"oster Schreiber, N. M., van der Wel, A., et al. 2011, \apj, 742, 96
\end{thebibliography}
\end{document}